\def\year{2019}\relax

\documentclass[conference]{IEEEtran}
\usepackage{cite}
\usepackage{amsmath,amssymb,amsfonts}
\usepackage{algorithmic}
\usepackage{graphicx}
\usepackage{url}
\usepackage{textcomp}
\usepackage{todo}
\usepackage{xcolor}

\usepackage{multirow}
\newcommand{\specialcell}[2][c]{%
  \begin{tabular}[#1]{@{}c@{}}#2\end{tabular}}
\newcommand{\framework}{\texttt{RGNet}}
\newcommand{\embedding}{\texttt{GUVec}}
\begin{document}

\title{Modeling Engagement Dynamics of Online Discussions using Relativistic Gravitational Theory}

\author{\IEEEauthorblockN{1\textsuperscript{st} Subhabrata Dutta}
\IEEEauthorblockA{
\textit{Jadavpur University}\\
Kolkata, India \\
subha0009@gmail.com }
\and
\IEEEauthorblockN{2\textsuperscript{nd} Dipankar Das}
\IEEEauthorblockA{
\textit{Jadavpur University}\\
Kolkata, India \\
dipankar.dipnil2005@gmail.com}
\and
\IEEEauthorblockN{3\textsuperscript{rd} Tanmoy Chakraborty}
\IEEEauthorblockA{
\textit{IIIT-Delhi, India}\\
Delhi, India \\
tanmoy@iiitd.ac.in}}
\maketitle

\begin{abstract}
  Online discussions are valuable resources to study user behaviour on a diverse set of topics. Unlike previous studies which model a discussion in a  static manner, in the present study, we model it as a time-varying process and solve two inter-related problems -- predict which user groups will get engaged with an ongoing discussion, and forecast the growth rate of a discussion in terms of the number of comments.  
  We propose \framework\ ({\bf R}elativistic {\bf G}ravitational {\bf Net}work), a novel algorithm that uses Einstein Field Equations of gravity to model online discussions as `cloud of dust' hovering over a user spacetime manifold, attracting users of different groups at different rates over time. We also propose \embedding, a global user  embedding method 
  for an online discussion, which is used by \framework\ 
  to predict temporal user engagement. \framework\ leverages different textual and network-based features to learn the dust distribution for discussions. 
  
  We employ four baselines -- first two using LSTM architecture, third one using Newtonian model of gravity, and fourth one using a logistic regression adopted from a previous work on engagement prediction. Experiments on Reddit dataset show that \framework\ achieves $0.72$ Micro F1 score and $6.01\%$ average error for temporal engagement prediction of user groups and growth rate forecasting, respectively, outperforming all the baselines significantly. We further employ \framework\ to predict non-temporal engagement -- whether users will comment to a given post or not. \framework\ achieves $0.62$ AUC for this task, outperforming  existing baseline by $8.77\%$ AUC. 
\end{abstract}

\section{Introduction}
\label{sec:intro}
Emergence of social media has resulted in a large-scale, heterogeneous and dynamic space for the users to get engaged in different activities. 
Studying engagement patterns in such platforms has its own merit for multiple purposes: market researchers can identify their potential audience for advertising campaigns and lucrative strategies; political campaigners can develop wide-scale trend analysis of the mass on the effect of their propaganda, etc.

Engagement dynamics in social media has thus attracted wide attention over a decade. Past studies attempted to predict (i) which pair of users is more likely to get engaged with each other based on their history \cite{schantl2013utility, yuan2016will}, and (ii) which posts will engage more users \cite{rowe2014mining, noguti2016post}. 
All these studies tackled the engagement prediction problem in a {\em static manner} by considering the entire discussion as a whole, thus ignoring dynamic user engagement and the micro-dynamics controlling temporal growth. 
The growth rate of a discussion, i.e., how many comments are being posted per unit time, varies over time, so as the user engagement. As the discussion continues, it unfolds diverse topics and user interactions, thus attracting different types of users over time. If we imagine users located in different points on a multidimensional space and clustered based on their coherent activities, a discussion can then be intuitively thought of as a growing and moving cloud in that space, attracting different sets of users in varying rates over time. \textit{The aim of the present work is to model the \textbf{time-varying engagement dynamics} of users with ongoing discussion -- a completely novel problem without any previous work, to the best of our knowledge.}  We build a framework which jointly models two phenomena -- user engagement from different clusters of users, and the rate of growth of discussions over time.

Different discussions {\em attract} different users at different rates. Although an individual user may get repelled by a particular discussion, the idea of repulsion cannot be consistently modeled without access to his/her cognitive data, or some platform-specific  features such as {\em dislike}. This implies that the interaction between a user and a discussion is essentially {\em attraction}, which can be zero but always non-negative. This motivates us to imagine a discussion to induce a gravity-like force towards the users. In fact, if we rely on the relativistic definition of gravity (explained in Sec. \ref{subsec:EFE}), it is even possible to adapt repulsion as a positive curvature in user manifold; however, in this work, we restrict ourselves to model interaction as `attraction' only. 

Newtonian model of gravitation describes gravity as a force following inverse squared distance law between particles -- proportional to the mass of the particles and inversely proportional to their distance squared. Given two point particles of mass $m_1$ and $m_2$ placed at positions $\Vec{r_1}$ and $\Vec{r_2}$ respectively, the magnitude of the force of gravity between them, denoted by $F$ is given by,
\begin{equation}
    \label{eq:newton}
    F = \frac{Gm_1m_2}{|\Vec{r_1}-\Vec{r_2}|^2}
    \vspace{-2mm}
\end{equation}
where $G$ is the gravitational constant. In our hypothesis, discussions have some mass-like property which changes over time. Users `near' to a discussion get attracted more. A  `massive' discussion tends to attract more users and therefore  would achieve more  growth rate. The degree of this `massiveness' can be a function of the topic, relevance, properties of engaged users, etc. But Newtonian model does not explain how mass and distance (or spacetime, to be precise) interact with each other. In case of online discussions, users are not mere objects, rather they have  histories, which bear complex connection with each other and the discussion itself. 

In physics, cosmic phenomena such as motion of the Mercury around the Sun \cite{le1859lettre}, bending of light passing near stars \cite{soldner1804deflection}, 
etc. cannot be explained by Newtonian model of gravity. A more sophisticated understanding of gravitation, which explains the failure of Newtonian model, was given by Einstein with his `General Theory of Relativity' \cite{einstein1915feldgleichungen} (GR Theory). Intuitively, relativistic theory of gravitation describes spacetime as an $(n+1)$-dimensional Riemannian manifold, with $n$ dimensions for space and one dimension for time. Gravity is simply the \textit{curvature} of this manifold at any point. According to this theory, the curvature can be caused by an object with mass and/or energy. 
Any object free-falling through this spacetime manifold must follow the `straightest' path or \textit{geodesic} -- a path with constant directional derivative w.r.t. the manifold. More the mass/energy content of an object, more curved the spacetime will be around it, and hence more will be the effect of gravity. {\bf This `fusion' of seemingly heterogeneous physical properties like mass/energy and  spacetime by GR theory is the primary motivation behind our proposed model \framework, which learns to efficiently fuse textual features of discussion with activity history of users in a temporal fashion to predict engagement dynamics.}

\begin{figure}[!t]
\centering
\includegraphics[width=0.45\textwidth]{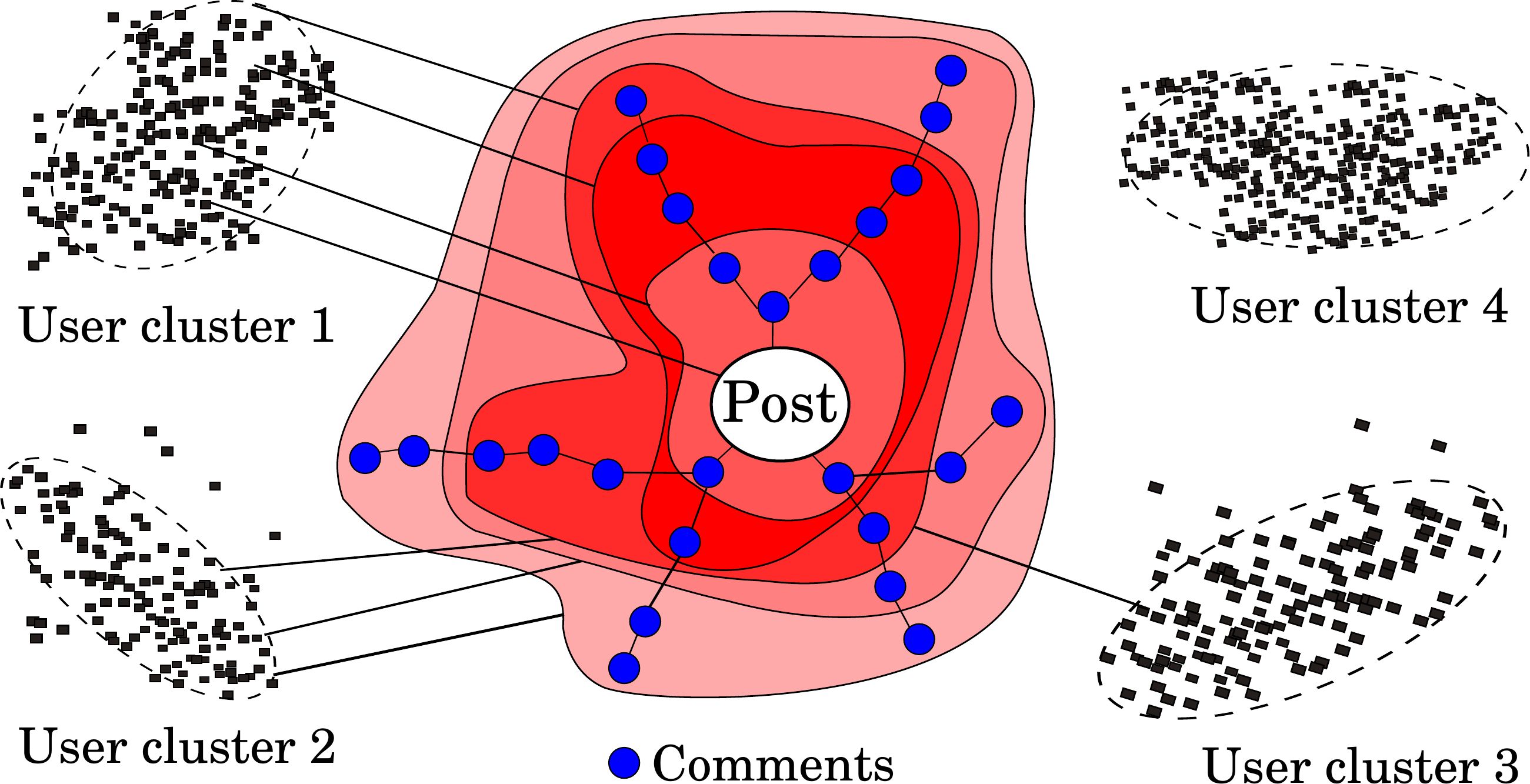}
\vspace{-2mm}
\caption{(Color online) Problem of temporal engagement dynamics explained: comments and their replies form a tree, with the post as root; each contour in the figure depicts a comment window of size 5; contours grow temporally; color intensity of the contour corresponds to the rate of growth, i.e.,  how fast the comments are added in that window; lines from the user cluster to contour indicate that users from that cluster commented on that window.}
\label{fig:model}
\vspace{-5mm}
\end{figure}

Fig. \ref{fig:model} explains how we model engagement dynamics as a time-varying process (Table \ref{tab:notation} summarizes important notations). In particular,
our major contributions are five-fold:
\begin{itemize}
    \item  We propose \embedding, a novel algorithm to represent users of a discussion platform as fixed dimensional vectors based on their temporal, communicative and semantic proximity.
     \item We propose \framework, an engagement prediction model which represents ongoing discussions as time-varying `dust clouds' in the user manifold 
     and models them using relativistic theory of gravity to predict which clusters of users from the manifold are likely to get engaged,
     and how fast the discussion cloud will grow.
     \item We propose two deep learning based models using Long Short-Term Memory (LSTM) cells, and a model similar to \framework\ based on Newtonian model of gravity. We also adopt the work by Rowe and Alani \cite{rowe2014mining} in temporal setting. All these models are considered as baselines.
     \item We  also predict user engagement by adopting \framework\ in a non-temporal setting (for the sake of a direct comparison with the existing baseline)   -- given a post, whether any user will comment to that post or not.
     \item We perform comprehensive evaluation on the Reddit CMV dataset \cite{tan2016winning} (for temporal engagement prediction) and Reddit \textit{r/news} community (for non-temporal engagement prediction) to show the efficiency of \embedding\ and \framework. 
     
\end{itemize}
     
     {\bf To the best of our knowledge, \framework\ is the first model of its kind which is inspired by the fundamental theories of classical mechanics.}

\begin{table}
    \centering
    \small
    \caption{Important notations used throughout the paper.}
    \vspace{-2mm}
    \label{tab:notation}
    \begin{tabular}{c|c}
    \hline
         {\bf Notation} & {\bf Denotation}  \\
         \hline
         $\mathbf{A}$ & User co-occurrence matrix\\
         $g_{\mu\nu}, g^{\mu\nu}$ & Metric tensor, inverse metric tensor\\
         $T_{\mu\nu}$ & Stress-energy tensor\\
         $R_{\mu\nu}, R$ & Ricci tensor, Ricci scalar\\
         $\sigma_1(x)$ & $max(0,x)$\\
         $\sigma_2(x)$ & $(1+\exp(-x))^{-1}$\\
         $w$ & Window size of comments\\
         $n$ & No. of user clusters \\
         $N$ & No. of windows in a discussion\\
         $\mathbf{U}$ & User set\\
         $C_i, \mathbf{C}$ & $i^{th}$ user cluster, set of cluster centers\\
         \hline
    \end{tabular}
    \vspace{-5mm}
\end{table}

 \section{\embedding: Global User Embedding}
 \label{sec:user_embedding}
 
 To compute user vectors from a discussion corpus, our proposed global user embedding method \embedding\ first constructs a user-user co-occurrence matrix $\mathbf{A}$. We use three different notions of proximity between two users $u_i$ and $u_j$: (i) {\bf Communicative Proximity:} they communicated with each other in a discussion; this happens when $u_i$ replied to $u_j$ in a discussion or vice versa, meaning they are present in the same chain of comments; (ii) {\bf Temporal Proximity:} they are temporally close to each other; they are engaged in same discussion (have not replied to each other) nearly at the same time; (iii) {\bf Semantic Proximity:} they are engaged in similar type of discussions.

 To construct a meaningful embedding of users, we only take those who are engaged in at least two discussions.
 Given the entire set of such users denoted by $\mathbf{U}$, the co-occurrence matrix $\mathbf{A}$ is symmetric and of dimension $|\mathbf{U}|\times|\mathbf{U}|$. To compute semantic proximity, we use ConceptNet Numberbatch word-vectors
 \cite{speer2016ensemble}. We take the words present in the discussion titles (after removing stopwords) and compute the weighted average of the corresponding word vectors. This weighted average now represents the title vector $T_k$ of  discussion $D_k$.

For any pair of users $u_i, u_j \in \mathbf{U}$, their proximity $A_{ij} = A_{ji} \in \mathbf{A}$ is computed as follows:
\begin{itemize}
     \item  Communicative Proximity: If $u_i$, $u_j$ replied to each other, then increment $A_{ij}$ by 2.
     \item Temporal Proximity: If $u_i$, $u_j$ commented on the same discussion at time $t_i$ and $t_j$ respectively, but did not reply to each other, then
     \begin{equation}
     \label{Eq:temp_sim}\small
     A_{ij}=A_{ij}+(1+\exp(-\alpha))^{-1}, \text{ where } 
         \alpha = \frac{t_{end}-t_{start}+1}{|t_i-t_j|+1}
     \end{equation}
     where $t_{start}$ and $t_{end}$ are the starting and ending times of the discussion, respectively.
     \item Semantic Proximity: If $u_i$, $u_j$  commented on different discussions $D_m$ and $D_n$, respectively, then 
    \begin{equation}
     \label{Eq:sem_sim}\small
         A_{ij} =
  \begin{cases}
                                   A_{ij}+\cos{\theta}& \text{if } \theta\leq\theta_0 \\
                                   A_{ij} & \text{otherwise}
  \end{cases}
  \vspace{-2mm}
\end{equation}
where $\theta = \arccos(\frac{T_m^\top  T_n}{|T_m|\cdot|T_n|})$, and  $0\leq\theta_0\leq\frac{\pi}{12}$ is a threshold angle (Sec. \ref{sec:setup} for parameter selection).

\end{itemize}
     
  In Eq.~\ref{Eq:temp_sim}, $\alpha$ accounts for how much temporally close two comments are w.r.t. the total time span of the discussion. This normalizes temporal proximity of discussions growing in different rates. We put highest proximity value for two users if they replied to each other. Both the terms $(1+\exp(-\alpha))^{-1}$ and $\cos\theta$ have upper bound of $1$. Therefore, for any pair of users, the contribution of their temporal and semantic proximity taken together cannot exceed their communicative proximity, which is incremented by 2.

Once $\mathbf{A}$ is computed, \embedding\ minimizes the following objective function
to obtain user vectors:
 \begin{equation}
 \small
 \label{Eq:embedding_loss}
 \vspace{-1mm}
     J = \sum_{i,j}^{|\mathbf{U}|} \log(1+A_{ij})(\mathbf{v_i}^\top \mathbf{v_j} +b_i+b_j -\log(1+A_{ij}))^2
     \vspace{-2mm}
 \end{equation}
 where $v_i$ and $b_i$ correspond to the $i^{th}$ user vector and bias, respectively. This objective function bears some similarity to that of GloVe embedding \cite{pennington2014glove}. Eq.~\ref{Eq:embedding_loss} uses the hypothesis that, for any two users $i$ and $j$, the term $\mathbf{v_i}^\top \mathbf{v_j}$ should be proportional to the logarithm of the probability of $j$ occurring in the context of $i$. This probability can be computed as $P_{ij}=A_{ij}/A_{i}$, thus $\log P_{ij}=\log A_{ij}-\log A_i$. Since $P_{ij}=P_{ji}$, i.e., the probability of a user $i$ appearing in user $j$'s context is same as the reverse, we need to exclude the term containing $A_{i}$. Hence we introduce the bias terms $b_i$ and $b_j$ in Eq.~\ref{Eq:embedding_loss}. We also need to assure that vectors of highly co-occurring users should be computed with greater accuracy. Therefore,  we introduce the weighing term $\log(1+A_{ij})$.

 \begin{figure}[!t]
\centering
\includegraphics[width=0.48\textwidth]{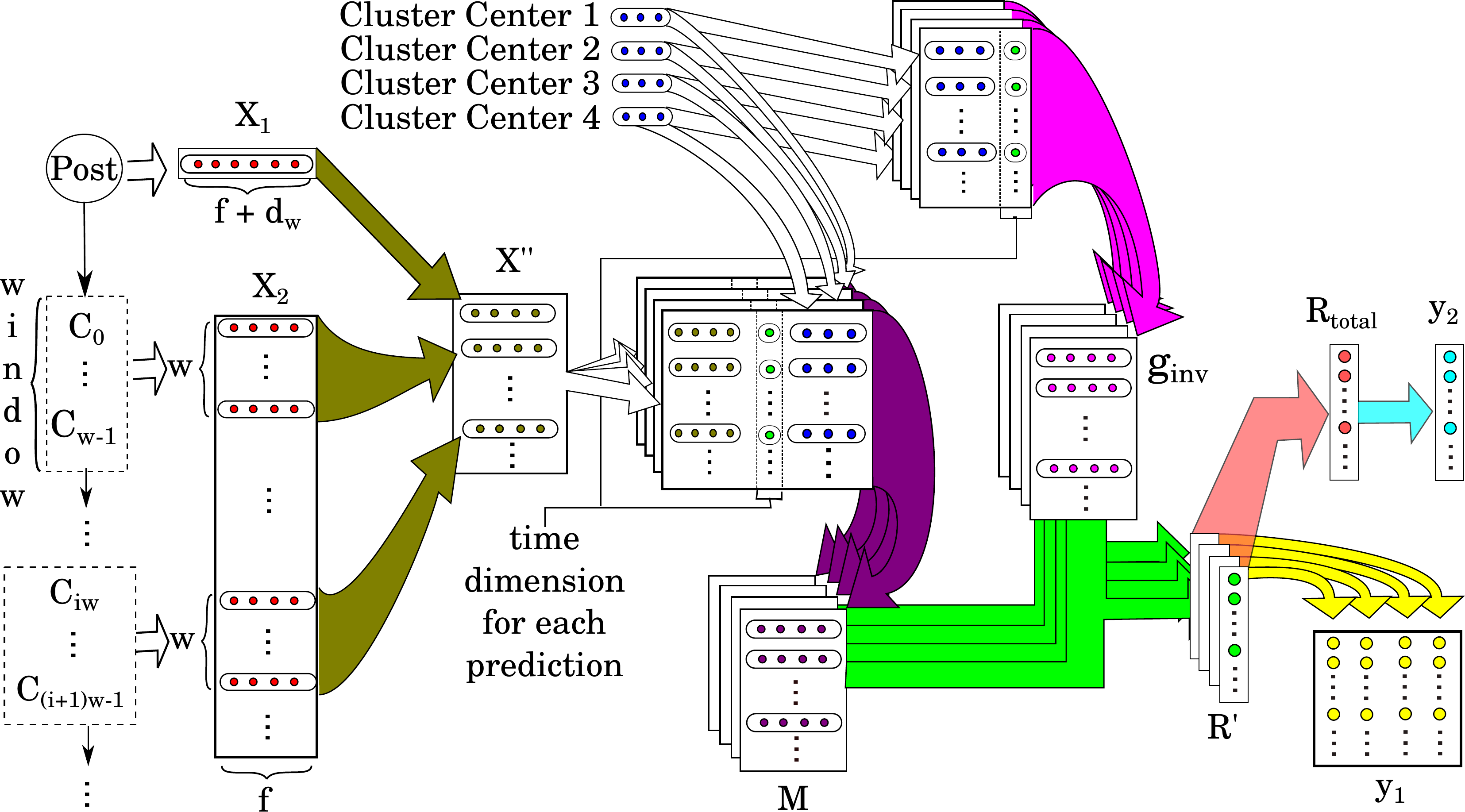}
\vspace{-3mm}
\caption{Architecture of \framework: C$_j$ signifies $j^{th}$ comment (chronologically ordered by time) in the discussion. Colored arrows represent different transformations as defined in Eq.~\ref{Eq:model_1} to Eq.~\ref{Eq:model_output}.}
\label{fig:rgnet}
\vspace{-5mm}
\end{figure}

 \section{\framework: Modeling User Engagement}
 After computing user vectors, we group them into $n$ clusters using standard clustering methods (see Sec. \ref{sec:setup}). Henceforth, the cluster centers $C_1,\cdots, C_n \in \mathbf{C}$ will represent $n$-regions of user manifold. We will first explain the Einstein Field Equations and their components, followed by how \framework\ incorporates them in modeling user engagement. Fig. \ref{fig:rgnet} shows a schematic architecture of \framework.
 
 \subsection{Einstein Field Equations (EFE)}
 \label{subsec:EFE}
 In general theory of relativity, spacetime is a four-dimensional manifold $\mathcal{M}$ with one dimension of time and three dimensions of space. Gravity is not an external force (like electromagnetic or nuclear forces), rather an intrinsic property of spacetime, defined as \textit{curvature} in $\mathcal{M}$. Any object without the effect of any force, will follow a {\em geodesic} (a curve for which directional co-variant derivative along the tangents of the curve remains zero) along this manifold. The geometry of the spacetime manifold is defined by sixteen Einstein Field Equations \cite{einstein1915feldgleichungen}:
 \begin{equation}
 \small
 \label{Eq.EFE}
     R_{\mu\nu}-\frac{1}{2}Rg_{\mu\nu}+\Lambda g_{\mu\nu} =\frac{8\pi G}{c^4}T_{\mu\nu}
 \end{equation}
 This is a tensor equation, with $\mu$, $\nu$ corresponding to dimension indices of the spacetime. As there are total four dimensions (one for time and three for space), the pair $\mu,\nu$ can take sixteen different values. $G$, $\Lambda$ and $c$ are three constants -- Newtonian gravitational constant, cosmological constant and velocity of light in vacuum, respectively. $g_{\mu\nu}$ is called the \textit{metric tensor} of the manifold. This is a {\em contra-variant} tensor which gives the idea of distance between two vectors on a manifold:
 \begin{equation}
 \small
 \label{Eq:metric_tensor}
     ds^2 = \sum_{\mu}\sum_{\nu}g_{\mu\nu}dx^{\mu}dx^{\nu}
     \vspace{-2mm}
 \end{equation}
 where $dx^i$ is the difference in the $i^{th}$ component of two vectors. It has its covariant counterpart $g^{\mu\nu}$, which is  called the {\em inverse metric}. 

$R_{\mu\nu}$ is the \textit{Ricci curvature tensor}. $R$ is the corresponding \textit{Ricci scalar}. The change in a vector for parallel transport (i.e., following a geodesic) along two different infinitesimal flows in a smooth manifold is given by the Riemann Curvature tensor. Ricci tensor is the contraction of Riemann tensor on the second index. 
Both Ricci and Riemann tensors can be computed from second order derivatives of the metric tensor. We define {\em Christoffel symbol of second kind} as $\Gamma_{kl}^{i}$,
\vspace{-1mm}
 \begin{equation}
 \small
     \label{eq:christoffel}
     \Gamma_{kl}^{i} = \frac{1}{2}g^{im}(g_{mk,l}+ g_{ml,k}-g_{kl,m})
 \end{equation}
 where $g_{mk,l}$ is the partial derivative of $g_{mk}$ with respect to the $l^{th}$ component. Then, $R_{ij}$ is defined as,
 \begin{equation}
     \label{eq:ricci_christoffel}
     R_{ij} = \Gamma_{ij,l}^{l}-\Gamma_{il,j}^{l}+ \Gamma_{ij}^{m}\Gamma_{lm}^{l} - \Gamma_{il}^{m}\Gamma_{jm}^{l}
 \end{equation}
 Eqs.~\ref{eq:christoffel} and \ref{eq:ricci_christoffel} are seemingly very complex to directly compute using metric tensor. However, the important fact is that  Ricci tensor can be computed as a function of derivatives of the metric tensor, and therefore, as differential function of the components of points in the manifold.
 Ricci scalar is simply the trace of the Ricci tensor:
 \vspace{-1mm}
 \begin{equation*}
 \small
     R = g^{\mu\nu}R_{\mu\nu}
     \vspace{-1mm}
 \end{equation*}
 Intuitively, in a 2D manifold, a zero Ricci scalar at a point indicates that the manifold is flat at that point; negative value indicates a saddle point, and positive value indicates a hill. In Eq.~\ref{Eq.EFE}, the term $R_{\mu\nu}-\frac{1}{2}Rg_{\mu\nu}$ describes the curvature of the spacetime manifold at any point. Its trace with respect to the inverse metric yields negative scalar curvature:
 \vspace{-1mm}
 \begin{equation}
 \small
     g^{\mu\nu}(R_{\mu\nu}-\frac{1}{2}Rg_{\mu\nu}) = -R
     \vspace{-1mm}
 \end{equation}
 \begin{figure}[!t]
\centering
\includegraphics[width=0.17\textwidth]{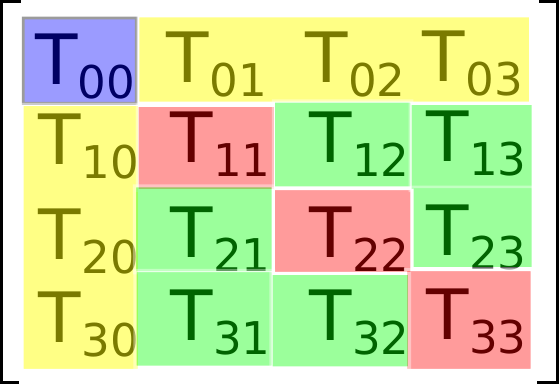}
\vspace{-2mm}
\caption{Components of the stress-energy tensor. The components in blue, yellow, green and pink represent mass density, momentum density, momentum flux and  pressure, respectively.}
\label{fig:stress-energy-tensor}
\vspace{-5mm}
\end{figure}
 $T_{\mu\nu}$ is called the \textit{stress-energy tensor}. For an infinitesimal volume of spacetime manifold, its components represent the properties as shown in Fig.~\ref{fig:stress-energy-tensor}. For an isolated massive particle, all the components except $T_{00}$ are zero. For a cloud of dust, only the diagonal elements have non-zero value.

Multiplying both sides of Eq.~\ref{Eq.EFE} by inverse metric tensor $g^{\mu\nu}$ yields,
 \begin{equation}
 \small
 \label{Eq:rel_curvature}
     \frac{8\pi G}{c^4}g^{\mu\nu}T_{\mu\nu} = -R+g^{\mu\nu}\Lambda g_{\mu\nu}
 \end{equation}
 
 \subsection{EFE in Discussion Spacetime}
Eq.~\ref{Eq.EFE} does not have any static solution without the cosmological constant $\Lambda$, indicating the universe is expanding. Einstein introduced $\Lambda$ to make it static, which, upon the observation of expanding universe in reality by Hubble \cite{hubble1929relation}, was discarded later. In our particular case of learning engagement dynamics in discussions using general relativity, we also omit  $\Lambda$ and reduce the constants in Eq.~\ref{Eq:rel_curvature} to yield
 \begin{equation}\small
 \label{Eq.model}
     \sum_{j=0}^{d}\sum_{k=0}^{d}g^{jk}(C_i)M_{jk}(C_i, T_i) = R'(C_i)
 \end{equation}
 where $C_i$ represents the position of $i^{th}$ user cluster in the user manifold we computed in Sec.~\ref{sec:user_embedding}, $T_i$ is the set of features representing the discussion, $d$ is the dimension of the user vectors, $g^{jk}$ is the inverse metric tensor which is computed as a function of cluster positions, $M_{jk}$ is the stress-energy tensor counterpart for discussion which is computed as a function of cluster positions and features of discussion. We prepend the time value to each user vector to convert it into a $(d+1)$-dimensional spacetime manifold. \framework\ learns each component of Eq.~\ref{Eq.model} as a series of non-linear transformations: $\mathbf{X'}=\sigma(\mathbf{W}^\top \cdot \mathbf{X} + \mathbf{B})$, where $\mathbf{X}$ and $\mathbf{X'}$ are input and output of the transformation respectively, $\sigma$ is a bounded non-linear function, $\mathbf{W}$ and $\mathbf{B}$ are weight and bias matrices to be learned respectively.

It is important to note that relativistic model of spacetime requires multiple constraints to be fulfilled. First of all, physical laws should be observer independent -- one can choose any frame of reference (rotated, translated, moving w.r.t. another frame of reference) and the physics must remain the same. General Relativity requires this constraint to be local. The user manifold obtained from \embedding\ computes the position of a user in the manifold using the vector dot product, which is invariant to rotation and translation. Moreover, it takes into account the temporal proximity of two users. We expect this to reflect invariance to temporal transformations of the manifold as well. However, engagement over online discussions is not a deterministic physical process. We claim it to be only analogous to spacetime geometry and not an exact replica. So in our case, Einstein's Field Equations are only abstract approximation learnt by \framework. An exact mathematical model of engagement is far more complex, if not intractable.

We define the temporal progress of a discussion as windows of comments of fixed size $w$. This means, at the $i^{th}$ step, the size of our discussion is $1+iw$ (post + comments), and we need to predict for the next $w$ comments. Due to variable size of discussions, we define maximum size of the discussion to be $(1+Nw)$, where $N$ is the number of windows, and hence, the number of prediction steps for a single discussion. All the discussions with size less than the maximum size are zero-padded at the end.

\subsection{Feature Selection}\label{sec:feature_selection}

For the original post and every comment in the discussion, we extract  following features based on the content, user, surface structure of the text.
 
\noindent{\bf (i)} {\bf Content Features:} \\
\indent{$\bullet$} {\em Average of tf-idf scores of the tokens}. This represents how many unique and relevant words are used in the comment.\\
\indent{$\bullet$} {\em LIX readability score} \cite{bjornsson1983readability}, computed as: 
         $r = \frac{|w|}{|s|}+100\times\frac{|cw|}{|w|}$,
where $w$ and $s$ are the sets of words and sentences respectively, and $cw$ is the set of words with more than six characters. Larger the value of $r$, harder the comment/post is to read in a short time.\\
\indent{$\bullet$} {\em Cumulative entropy of terms}, given by $
         p=\frac{1}{|T|}\sum_{t\in T}tf_t(\log|T|-\log(tf_t))$,
where $T$ is the set of all unique tokens in the corpus, and $tf_t$ is the frequency of term $t$ in the comment/post.\\
\indent{$\bullet$} {\em Polarity of the comment/post}, i.e., sum of sentiment intensity scores of the unique terms computed using SenticNet \cite{cambria2018senticnet}. We also use the total number of positive and negative sentiment words as polarity features.

\noindent{\bf (ii)} {\bf Surface Features}: We use the following surface features -- total number of sentences in the text, average number of words per sentences, count of URLs present in the comment, depth of the comment in discussion tree, time difference of the comment with the post and the count of closing punctuation markers, i.e., `.',`!' and `?' (as different types of closing punctuation markers signify different discourse).

\noindent{\bf (iii)} {\bf Latent Semantics}: We use the pre-trained word vectors mentioned in Sec. \ref{sec:user_embedding} to represent the latent semantics of the text. Every comment is represented as a vector: $
         V = \frac{1}{|C|}\sum_{t\in C} (tf\text{-} idf_t\cdot W_t$),
where $C$ is the set of unique terms in the comment, and $W_t$ is the word vector of term.
For the post, we also use the title vectors mentioned in Sec. \ref{sec:user_embedding} as features. 

\noindent{\bf (iv) User Features}: We use user vectors computed by \embedding\ as user-based features, which reflect past activity and connections of a user.

For a total $f$ number of features representing each comment, the representation of a post $\mathbf{X_1}$ is then an array of size $(f+d_w)$ with $d_w$ being the size of word vectors used; all the comments taken together $\mathbf{X_2}$ is an array of size $N\times w\times f$, and user manifold regions $\mathbf{C}$ are represented as an array of size $N\times n\times (d+1)$ for a single discussion.

 \subsection{Stress-Energy Tensor of Discussion}
 First, we compute an intermediate representation of the post and the comments with dimension:
 \vspace{-1mm}
 \begin{equation}
 \small
 \label{Eq:model_1}
 \begin{split}
     \mathbf{X'_1} &= \sigma_1(\mathbf{W_1}\cdot \mathbf{X_1}^\top + \mathbf{B_1})\\
     \mathbf{X'_2}[i] &= \sigma_1(\mathbf{W_2}\cdot \mathbf{X_2}[i]^\top + \mathbf{B_1})\\
     \mathbf{X'} &= (\mathbf{X'_1}, \mathbf{X'_2})
 \end{split}
 \vspace{-1mm}
 \end{equation}
$\mathbf{X'}$ now contains representation of the post and each of the $N$ comment windows. $\mathbf{W_k}$ and $\mathbf{B_k}$ (where $k=1,2,\cdots$) mentioned throughout the paper indicate the learnable weight and bias matrices, respectively.
$\sigma_1$ is the rectified linear unit function. Now all the representations from 0 to $(i-1)^{th}$ steps should contribute at  $i^{th}$ step. Therefore, we take a weighted cumulative sum of $\mathbf{X'}$:
\begin{equation}
\small
 \label{Eq:model_2}
 \mathbf{X''}[i]= \frac{\sum_{j=0}^{i}\mathbf{W_3}[j]\mathbf{X'}[j]}{\sum_{j=0}^{i}w_j}
 \end{equation}
 
 Next, we concatenate $\mathbf{X''}[i]$ to each $C_l\in \mathbf{C}$ and compute the corresponding stress-energy tensor:
 \begin{equation}
 \small
 \label{Eq:model_3}
 \begin{split}
 \mathbf{M}[i][l] &= \sigma_2(\mathbf{W_4}\cdot\sigma_2(\mathbf{W_5}\cdot (\mathbf{X''}[i],C_l)^\top\\
 &+\mathbf{B_4})^\top+\mathbf{B_3})
 \end{split}
 \end{equation}
 where $\sigma_2(x)=(1+\exp(-x))^{-1}$. Each $\mathbf{M}[i][l]$ is a $(d+1)$-dimensional vector representing the diagonal elements of the stress-energy tensor $M_{jk}$ of Eq.~\ref{Eq.model}. 
 
\subsection{Inverse  Metric Tensor}
 We compute the values of inverse metric tensor $\mathbf{g}_{inv}$ for $i^{th}$ prediction step at $l^{th}$ cluster region as a function of the cluster center as follows:
 \vspace{-1mm}
 \begin{equation}
 \small
 \mathbf{g}_{inv}[i][l] = \sigma_2(\mathbf{W_6}\cdot\sigma_2(\mathbf{W_7}\cdot C_l^\top+\mathbf{B_6})^\top +\mathbf{B_5})
 \vspace{-1mm}
 \end{equation}
 Again, this is a $(d+1)$-dimensional vector which represents the diagonal of the inverse metric tensor $g^{jk}$ of Eq.~\ref{Eq.model}.
 
 \subsection{Curvatures of Manifold}
Once we obtain the stress-energy tensor $\mathbf{M}[i][l]$ and the inverse matrix tensor $\mathbf{g}_{inv}[i][l]$ of the discussion at $i^{th}$ prediction step for $l^{th}$ cluster, we  compute the scalar curvature of the manifold at $l^{th}$ cluster based on Eq. \ref{Eq.model} as,
\vspace{-1mm}
 \begin{equation}
 \small
     R'[l] = \sum_{j=0}^{d}\mathbf{M}[i][l][j]\cdot\mathbf{g}_{inv}[i][l][j]
     \vspace{-1mm}
 \end{equation}
 This step actually performs the fusion of textual and user interaction features. The extent to which the discussion attracts users towards it for the entire manifold can be computed as the weighted sum of each of $R'[l]$, given by,
 \vspace{-1mm}
 \begin{equation}
 \label{eq:r_total}
 \small
     R_{total} = \sum_{l=1}^{n}\mathbf{W_8}[l]\cdot R'[l]
     \vspace{-1mm}
 \end{equation}
 Finally, we define cluster engagement probability $y_1$ and discussion growth velocity $y_2$ as two nonlinear functions of cluster-wise scalar curvature and total curvature respectively:
 \vspace{-1mm}
 \begin{equation}
 \small
 \label{Eq:model_output}
         y_1 = \sigma_2(R'),\ \ \ 
         y_2 = \sigma_1(R_{total})
         \vspace{-1mm}
 \end{equation}
 so that $0\leq y_1\leq 1$ and $0\leq y_2$, befitting to both user cluster engagement prediction and growth rate forecasting tasks. Altogether, we train \framework\ to learn the following function:
 \vspace{-2mm}
 \begin{equation}
 \small
     (y_1,y_2)=\mathcal{F}(\mathbf{X_1},\mathbf{X_2},\mathbf{C}|\mathbf{W_{1,\cdots 8}}, \mathbf{B_{1,\cdots 6}}) 
     \vspace{-1mm}
 \end{equation}

For user cluster engagement prediction task (multi-label classification), we use \textit{binary cross-entropy} loss with $0.5$ as the threshold, and for the growth rate forecasting (regression),  we use \textit{mean squared error} loss to train \framework. 

 \section{Baselines for Temporal Engagement}
 \label{sec:baseline}
 Due to the lack of existing baseline in predicting temporal user engagement, we design four baselines:

 (i) {\bf Newtonian Model:} This model is similar  to \framework\ except it uses  Newtonian model over flat space instead of spacetime manifold. We compute $\mathbf{X''}$ based on Eqs. \ref{Eq:model_1} and \ref{Eq:model_2} using same set of features, except instead of computing a stress-energy tensor for each cluster (Eq. \ref{Eq:model_3}), we compute a scalar mass $M_i$ at the $i^{th}$ prediction step:
 \begin{equation}\small
 \label{Eq:New_model_3}
 \begin{split}
 M_i &= \sigma_2(\mathbf{W'_1}\cdot\sigma_2(\mathbf{W'_2}\cdot \mathbf{X''}[i]^\top +\mathbf{B'_1})^\top+\mathbf{B'_2})
 \end{split}
 \end{equation}
 
 We compute the position of the discussion at $i^{th}$ prediction step on the $d$-dimensional space as a weighted average of user vectors $\mathbf{r}$ commented till $i^{th}$ prediction,
 \begin{equation}
 \small
     r_i = \Big(\sum_{j=0}^{iw-1}\mathbf{W'_3}[j]\Big)^{-1} \Big(\sum_{j=0}^{iw-1}\mathbf{W'_3}[j]\cdot \mathbf{r}[j] \Big)
 \end{equation}
 
 We then define $y_1$ and $y_2$ from Eq.~\ref{Eq:model_output} as,
 \vspace{-1mm}
 \begin{equation}
 \small
 \label{Eq:newmodel_output}
 y_1 = \sigma_2(\frac{M_i}{|r_i-C_l|^2});\ \ y_2 = \sigma_1(\sum_{l=1}^{n}\mathbf{W'_4}[l]\cdot \frac{M_i}{|r_i-C_l|^2})
 \vspace{-1mm}
 \end{equation}
 and train the model using same loss functions.

\begin{figure}[!t]
\centering
\includegraphics[width=0.48\textwidth]{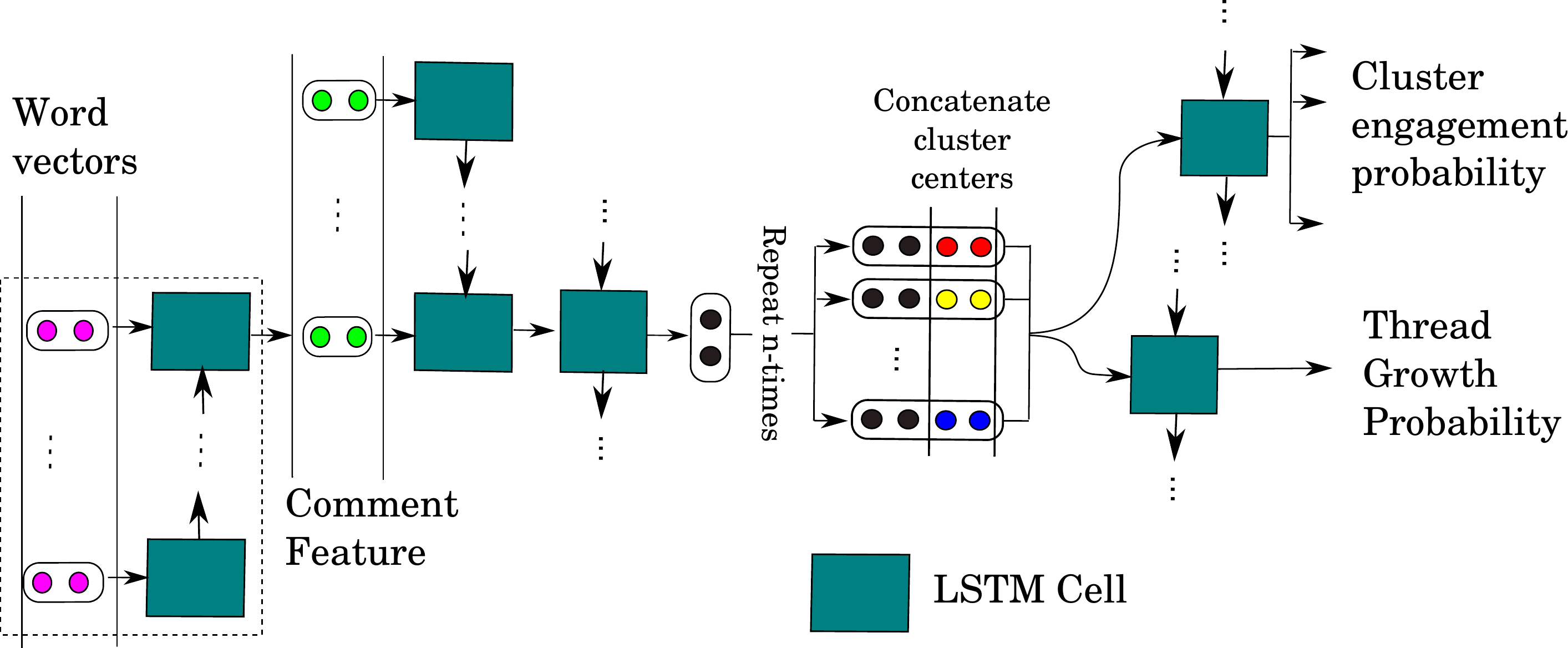}
\vspace{-2mm}
\caption{Architecture of the LSTM-f and LSTM-r models. LSTM layer in dotted region present in the left is used only for LSTM-r.}
\label{fig:rnn_baseline}
\vspace{-5mm}
\end{figure}

(ii) {\bf LSTM Models:} We implement two LSTM models; one with the features we defined in Sec. \ref{sec:feature_selection} ({\bf LSTM-f}), another using raw text data ({\bf LSTM-r}). To input raw text data, we use one-hot encoding of each word and initialize an embedding layer with pre-trained word vectors mentioned in Sec.~\ref{sec:user_embedding}. This model uses an extra layer of LSTM cells to compute the representation of comments from words. Both these models use same loss functions (binary cross-entrpoy and mean squared error). Fig. \ref{fig:rnn_baseline} shows the architecture.

(iii) {\bf Logistic Regression:} Lastly, we implement a logistic regression classifier adapted from the model proposed by Rowe and Alani \cite{rowe2014mining}. We consider the same set of features except {\em the duration of a user in the community} as Reddit does not provide this data. The authors broadly categorized the features used as \textit{social} and \textit{content} features. Their original work is not designed for temporal engagement modeling. Also, they performed a binary classification of whether a post will get commented or not. We adopt this model for our task with two modifications: (a) we take each user cluster and predict whether a user from this cluster will comment (user-network based features are calculated for each cluster separately, not the whole user-user interaction network) and (b) at each prediction step for a particular discussion, we take the post and comments (if any) till that instance as a single entity -- content features are calculated from the merged texts of post and comments, and the average of social features of the users who posted/commented is considered as the cumulative social feature. {\em This model is  made only for predicting engagement of user clusters}. 

\section{Non-temporal Engagement Prediction}
As already stated, our defined problem of predicting temporal engagement dynamics is novel, and there is no existing work which can be directly considered to compare with \framework. Rowe and Alani \cite{rowe2014mining} (henceforth, referred as R\&A) proposed a framework to predict engagement in a non-temporal manner. Given a post, their model predicts whether it will attract any user or not. We modify \framework\ to suit this task and compare the performance. 

We hypothesize that, if a post fails to curve the user manifold `effectively', it will not attract any users in the future. For this, we input the post feature $\mathbf{X_1}$ to \framework. As this is a one-shot prediction for the post only, the comment feature $\mathbf{X_2}$ as well as the comment window are irrelevant here. Also, all the occurrences of $i$ in the governing equations of \framework\ have a single value (i.e., $0$), as this is the first prediction step in the full implementation of \framework. Therefore, the stress-energy tensor $\mathbf{M}$ in Eq. \ref{Eq:model_3} is computed from $\mathbf{X'_1}$ only (first part of Eq. \ref{Eq:model_1}). Total curvature $R_{total}$ (in Eq. \ref{eq:r_total}) estimates the degree of total attraction generated by the post. We compute the probability of a post to attract any user at all from the total curvature as:
\vspace{-1mm}
\begin{equation}
\small
    y_3 = \sigma_2(R_{total})
    \vspace{-1mm}
\end{equation}
Here, $y_3$ ranges in interval $(0,1)$. We take $y_3\leq 0.5$ as negative class (post fails to attract any user), and positive class, otherwise.

\section{Experimental Setup}\label{sec:setup}
We describe the datasets and parameter selection for \framework: both for temporal and non-temporal engagement prediction.
\subsection{Datasets}
The Reddit CMV dataset that we used contains $18,363$ discussions from Jan 1, 2013 - May 7, 2015 for training, and $2,263$ discussions from May 8, 2015 - Sep 1, 2015 for testing. We excluded comments posted by deleted users and delta-bots (carrying author-tags ``deleted'' and ``DeltaBot'' respectively) and users who commented only once. 
This leaves our training (test) set with $46,121$ ($6,044$) users and $1,011,890$ ($112,432$) comments in total.
 
However, this CMV dataset was originally filtered, such that there is no post which failed to attract any user comment. Therefore, we cannot use this dataset for non-temporal engagement prediction. For this task, we crawled posts from Reddit \textit{news} community. We collected a total of $43,343$ posts from Sep 1, 2016 to Jan 16, 2019, out of which $5,449$ posts do not have any comments. To avoid classification bias, we take equal number of posts containing comments. Here again, we excluded users who have commented/posted only once or carry the author tag ``deleted'' (delta-bots does not appear in this community) to compute the \embedding\ embeddings. This results in a total of $29,431$ users.

\subsection{Parameter Selection}
\begin{figure*}[!t]
\centering
\includegraphics[width=0.8\textwidth]{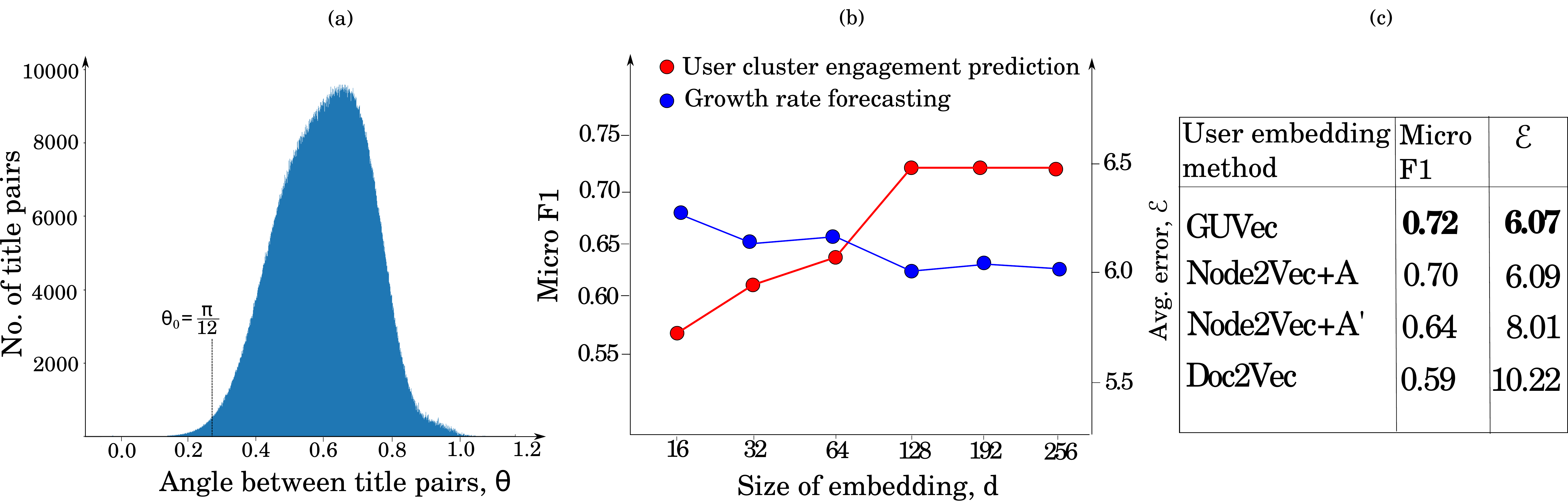}
\vspace{-2mm}
\caption{(a) Selection of $\theta_0$, (b) variation of \framework's performance w.r.t. $d$, (c) performance of user embedding methods for both the temporal tasks with \framework.}
\label{fig:parameter}
\vspace{-5mm}
\end{figure*}

While constructing the  co-occurrence matrix $\mathbf{A}$, computing the semantic proximity is computationally the most expensive part as we need to count for all possible pairs of users between every pair of discussions. The choice of  $\theta_0$ in Eq.~\ref{Eq:sem_sim} can significantly reduce this cost if we pre-compute  $\cos\theta$ between pairs of  discussion titles and take into account only those discussions having $\theta \leq \theta_0$. In Fig.~\ref{fig:parameter}(a), we plot the number of discussion title pairs with $\theta$ between them. Discussion pairs with $\theta\leq \frac{\pi}{12}$ amounts to $1.5\%$ of the total pairs. We find that the number of user-pairs for this subset is $\mathcal{O}(|\mathbf{U}|^{\frac{3}{2}})$. Therefore we choose $\theta_0$ to be $\frac{\pi}{12}$.

We vary the embedding dimension $d$ from 16 to 256.
Fig.~\ref{fig:parameter}(b) shows that the performance of \framework\ does not change much after $d=128$. 
We cluster user embedding space using K-means by varying K from $8$, $16$, $24$ to $32$.\footnote{We also tried with agglomerative and DBSCAN methods for user clustering and observed similar results.}
%
In case of engagement modeling, we vary window size from $5$ to $20$.  All the models except the logistic regression were optimized using Adam \cite{kingma2014adam} optimization algorithm. Unless otherwise stated, we use the following parameter values as default: $\theta_0 = \frac{\pi}{12}$, $d=128$, $w=15$, and $K=n=8$. 


\begin{table*}[!t]
     \centering
     \caption{Evaluation metrics used for multi-label classification. There are $n$ multi-label instances $(x_i, y_i)$ with $y_i$ being a binary vector of size $L$, and $h(x_i)$ being the predicted set of labels for $x_i$. $\lVert a\rVert_1$ denotes L1 norm of vector $a$, $\bigoplus$ signifies element-wise XOR, $y_k$ and $h^k(x_i)$ are $k^{th}$ entries of true label set and predicted label set respectively, and $I(s)=1$ if $s$ is true, and $0$ otherwise.}
     \small
     \begin{tabular}{|l|l|l|}
     \hline
          \multicolumn{1}{|c|}{\bf Metric} & \multicolumn{1}{c|}{\bf Description} & \multicolumn{1}{c|}{\bf Formula}  \\
     \hline
          Micro F1 & Micro average of precision and recall on all binary labels & $\frac{2\times \sum_{i=1}^{n}\lVert h(x_i)\cap y_i \rVert_1}{\sum_{i=1}^{n} \lVert h(x_i)\rVert_1+\sum_{i=1}^{n}\lVert y_i\rVert_1}$\\
    \hline
          Macro F1 & Macro average of precision and recall  
          & $\frac{1}{L}\sum_{k=1}^{L}\frac{2\times \sum_{i=1}^{n}h^k(x_i)y_i^k}{\sum_{i=1}^{n}h^k(x_i)+\sum_{i=1}^{n}y_i^k}$\\
     \hline
          Hamming Loss & Average error rate over all the binary labels
          & $\frac{1}{n}\sum_{i=1}^{n}\frac{1}{L}\lVert h(x_i)\bigoplus y_i\rVert_1$\\
     \hline
          Subset $0/1$ Loss & Average \% when predicted label set is exactly correct
          & $\frac{1}{n}\sum_{i=1}^{n}I(h(x_i)\neq y_i)$\\
     \hline
     \end{tabular}
     \vspace{-5mm}
     \label{tab:metrics}
 \end{table*}

\section{Experimental Results}

We perform comparative evaluation for three tasks separately. For temporal engagement dynamics, we compare \framework\ with other baselines to predict user cluster engagement and growth rate forecasting. For non-temporal engagement, we present the performance of \framework\ for different number of clusters and compare it with R\&A  \cite{rowe2014mining}. We study the importance of different features for these tasks. We also show the efficiency of \embedding\ compared to other embedding methods for temporal engagement tasks mentioned above, and empirically show the complexity of \embedding. In the end, we present a case study of user cluster engagement prediction obtained from \framework.

\begin{table}[]
    \centering
    \caption{Multi-label classification performance of the competing methods for temporal user engagement prediction. $\downarrow$ ($\uparrow$) indicates the smaller (larger) the value, the better the performance.}
    \scalebox{0.9}{
    \begin{tabular}{|l|c|c|c|c|}
    \hline
         {\bf Method} & {\bf HL} $\downarrow$ & {\bf MiF} $\uparrow$ & {\bf MaF} $\uparrow$ & {\bf 0/1} $\downarrow$   \\\hline
         \framework & {\bf 0.27} & {\bf 0.72} & {\bf 0.65} & {\bf 0.78} \\\hline
         LSTM-f & 0.36 & 0.64 & 0.57 & 0.81 \\
         LSTM-r & 0.37 & 0.61 & 0.56 & 0.81  \\
         Newton. & 0.40 & 0.56 & 0.51 & 0.86   \\
         Logistic Regression & 0.36 & 0.62 & 0.56 & 0.85 \\\hline
    \end{tabular}}
    \vspace{-5mm}
    \label{tab:accuracy_cluster}
\end{table}

\subsection{Predicting Temporal Engagement of User Clusters}
We pose user cluster engagement prediction problem as a multi-label classification problem. 
At $i^{th}$ prediction step, let there be $m$ comments in the $(i+1)^{th}$ window, with $m\leq w$. Let there be $n$ clusters of the user manifold. Each instance in our dataset corresponds to a window. For $(i+1)^{th}$ window, we  create the ground-truth binary vector $\mathbf{Y_{i+1}}$ of size $n$ such that, $Y_{i+1}[j]=1$ if there is at least one comment in the $(i+1)^{th}$ window from a user belonging to $j^{th}$ cluster, $0$ otherwise.

Table \ref{tab:accuracy_cluster} reports the performance of the competing methods based on four standard metrics used for multi-label classification \cite{Shi:2014:MCB:2611448.2505272} (see Table \ref{tab:metrics} for the description) -- Hamming Loss (HL), Micro F1 (MiF), Macro F1 (MaF), and Subset 0/1 (0/1). \framework\ outperforms others across all the metrics -- it beats the best baseline (LSTM-f) by 12.5\% (14.03\%) higher  Micro (Macro) F1 .

Table \ref{tab:varyingwindowcluster}(a) shows that as the number of clusters $n$ grows, the average degradation of performance for \framework\ and Newtonian model is minimum (15.22\% and 10.15\% respectively averaged across consecutive values of $n$) compared to others (15.14\% for LSTM-f).  These two models  benefit from the fact that, with smaller cluster size, cluster centers exhibit accurate locality of the cluster, which helps them compute more accurate curvature and the distance vector.  Table \ref{tab:varyingwindowcluster}(b) shows that for most of the models, the performance increases as the window size $w$ grows.

To check how homogeneity of users (in terms of their clusters) already engaged till $i^{th}$ window affects the performance for $(i+1)^{th}$ window, we compute the
entropy of the cluster membership of users $\mathbf{U}_{i}$ engaged till $i^{th}$ window, as $H_i=-\sum_{c\in \mathbf{C}} p_c \log p_c$, where $p_c$ is the fraction of users in $\mathbf{U}_{i}$ belonging to cluster $c$. Fig. \ref{fig:homogeneity}(a) indicates that as $H_i$ increases (users already engaged tend to be members of same cluster), the performance decreases (Pearson $\rho=-0.632$) since the model tends to predict more to the cluster whose members are engaged more in discussion. This further results more mistakes for those clusters which have not been engaged so far in the discussion. However, the decrease in performance  is less for \framework\ compared to the best baseline.

\begin{table}[]
\centering
\caption{Performance with different number of clusters and window size for both the temporal tasks.}

\begin{tabular}{|l|c|c|c|c|}
\hline
\multirow{2}{*}{\textbf{Model}} & \multicolumn{4}{c|}{{\bf (a) \# of clusters, $n$ (Micro F1, $\mathcal{E}$)}}\\\cline{2-5}
& 
\textbf{8} & \textbf{16} & \textbf{24} & \textbf{32}\\\hline
\framework & 0.72,11.02 & 0.62,7.93 & 0.53,6.54 & 0.44, 6.01 \\ 
LSTM-f & 0.64,11.23 & 0.57,10.87 & 0.46,9.92 & 0.39,10.56 \\
LSTM-r & 0.61,21.55 & 0.53,19.12 & 0.46,20.04 & 0.38,21.87 \\
Newt. & 0.56,24.66 & 0.48,28.54 & 0.42,25.01 & 0.40,23.12  \\\hline
\end{tabular} 

\begin{tabular}{|l|c|c|c|c|}
\hline
\multirow{2}{*}{\textbf{Model}} & \multicolumn{4}{c|}{{\bf (b) Window size, $w$ (Micro F1, $\mathcal{E}$)}}\\\cline{2-5}
& 
\textbf{5} & \textbf{10} & \textbf{15} & \textbf{20}\\\hline
\framework & 0.54,8.04 & 0.66,6.07 & 0.72,6.01 & 0.69,6.12 \\ 
LSTM-f & 0.49,12.56 & 0.55,11.20 & 0.64,11.23 & 0.65,10.98 \\
LSTM-r & 0.49,24.32 & 0.54,22.87 & 0.61,21.55 & 0.61,21.45 \\
Newt. & 0.32,29.57 & 0.46,27.01 & 0.56,24.66 & 0.57,25.03
\\\hline
\end{tabular} 
\label{tab:varyingwindowcluster}
\vspace{-5mm}

\end{table}

\subsection{Growth Rate Forecasting for Temporal Engagement}

We define the {\em growth rate} of engagement for a discussion  at $(i+1)^{th}$ window as $v^{i+1} = \log(\frac{m}{\Delta t})$, where $\Delta t$ is the time difference of first and last comments in $(i+1)^{th}$ window, and $m$ is the window size.
To test how effectively each competing model predicts the growth rate for $(i+1)^{th}$ window, we use relative \%-error in prediction given by $\mathcal{E}^{i+1}=\frac{|v^{i+1}_{true}-v^{i+1}_{predict}|}{|v^{i+1}_{true}|}\times100\%$, where $v^{i+1}_{true}$  ($v^{i+1}_{predict}$) is the actual (predicted) value. Table \ref{tab:varyingwindowcluster}(a) shows the average error $\mathcal{E}$ across all the windows incurred by the models trained with different number of clusters $n$. We observe that as $n$ increases, the average error for \framework\ decreases. The reason is that more the number of clusters obtained from user embedding, more precisely \framework\ can compute the curvature throughout the manifold. 

\begin{figure}[!t]
\centering
\includegraphics[width=\columnwidth]{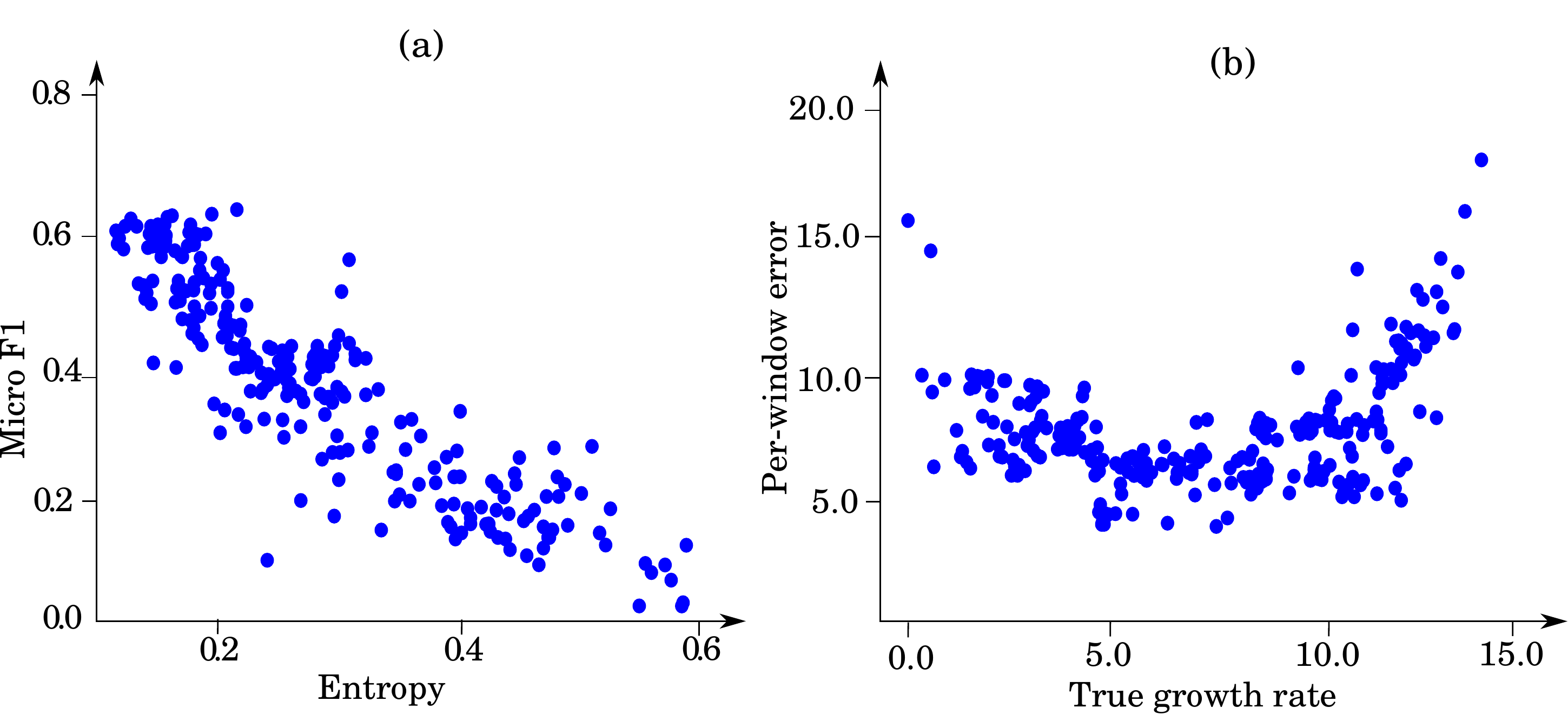}
\caption{(a) Accuracy of \framework\ with the change in homogeneity (measured by entropy), and (b) per-window error with increasing growth rate.}
\label{fig:homogeneity}
\vspace{-5mm}
\end{figure}

Fig.~\ref{fig:homogeneity}(b) shows the correlation of per-window error $\mathcal{E}^{i+1}$ and true growth rate $v^{i+1}_{true}$ (for better visualization, we normalize $v^{i+1}_{true}$ by its maximum value obtained). We observe that for higher values of growth rate, prediction is more erroneous (Pearson $\rho=0.178$). We empirically observe that such a large growth rate occurs when more that $50$ comments appear per second. Such instances (discussions) seldom appear in our dataset (1.74\% of total discussions).

\subsection{Non-temporal Engagement Prediction}
\begin{table}[!t]
    \centering
    \caption{Evaluation results for predicting non-temporal engagement. \framework-$n$ signifies \framework\ with $n$ number of clusters.}
    \vspace{-1mm}
    \begin{tabular}{|c|c|c|c|}
    \hline
    {\bf Model} & {\bf F1-score} & {\bf AUC} & {\bf Accuracy}\\
    \hline
         \framework-8 & 0.51 & 0.52 & 0.52 \\
         \framework-16 & 0.54 & 0.55 & 0.57 \\
         \framework-24 & {\bf 0.61} & {\bf 0.62} & {\bf 0.63} \\
         \framework-32 & 0.59 & 0.59 & 0.60 \\
         R\&A & 0.56 & 0.57 & 0.58\\\hline
    \end{tabular}
    \vspace{-5mm}
    \label{tab:non-temporal-engagement}
\end{table}
In Table \ref{tab:non-temporal-engagement}, we present the performances of different models for predicting whether a given post will attract users or not. We observe that \framework\ with 24 clustering of the manifold
performs the best. Moreover, both settings with 24 and 32 clusters outperform R\&A by a significant margin. 

 In Table \ref{tab:non-temporal-engagement}, we can observe an increase in performance of \framework\ for this task as the number of clusters grows. A possible reason  might be the heterogeneous distribution of users over the manifold and how accurately \framework\ is being informed about this heterogeneity. It is already explained that more closely located users (i.e., users in a dense cluster) are more likely to interact with each other in near future. Therefore, a post coming from an outlier user is less likely to be replied by other users. With less number of clusters, sparsely separated users are identified to be members of a cluster. This results in \framework\ assigning wrong curvature value for those users. With more clusters, this error is minimized. However, with increasing number of clusters, errors in curvature computation for each cluster get accumulated  and affect the total curvature $R_{total}$. This possibly explains the performance drop for \framework\ with $32$ clusters in Table \ref{tab:non-temporal-engagement}.

\subsection{Feature Importance}
We perform feature ablation study for both temporal and non-temporal engagement prediction tasks. For the former case, we study feature importance only for \framework, whereas for the latter case, the analysis is done for both \framework\ and R\&A.

\subsubsection{Feature Ablation for Temporal Engagement}
We study the importance of different features for \framework\ in two settings. In the first setting, 
we drop each group of features (mentioned in Sec. \ref{sec:feature_selection}) in isolation and report the accuracy. In the second setting, we add random noise to each feature in isolation --  we draw random samples from Gaussian distributions with same mean and standard deviation as that of the original distribution of the feature (this experiment was repeated 10 times, and the average result was reported). Table~\ref{tab:feature_study} indicates that user features bear utmost importance for both the tasks, though its effect is more visible in user cluster engagement prediction than the growth rate prediction. This is quite consistent with our intuition that {\em similar types of users tend to flock together}.

\begin{table}[]
\centering
\small
\caption{Performance of \framework with each feature set (i) removed in isolation, and (ii) replaced by noisy feature set in isolation for both user cluster prediction (Micro F1) and growth rate forecasting ($\mathcal{E}$). }
\vspace{-2mm}

\begin{tabular}{|c|c|c|c|c|}
\hline
\multirow{2}{*}{{\bf Feature set}} & \multicolumn{2}{c|}{\specialcell{{\bf Feature removed}}} & %
    \multicolumn{2}{c|}{\specialcell{{\bf Noise added}}}\\
\cline{2-5}
& {\bf Micro F1} & {\bf $\mathcal{E}$} & {\bf Micro F1} & {\bf $\mathcal{E}$}\\
\hline
Latent  & 0.62 & 8.33 & 0.57 & 16.56 \\ 
User & {\bf 0.54} & {\bf 8.34} & {\bf 0.52} & {\bf 17.15} \\ 
Content & 0.59 & 7.12 & 0.56 & 15.42  \\ 
Surface & 0.63 & 8.43 & 0.57 & 14.89 \\
\hline

\end{tabular}
\vspace{-5mm}
\label{tab:feature_study}
\end{table}
\begin{table}[!t]
\centering
\caption{Feature ablation of  \framework\ and R\&A for the task of non-temporal user engagement prediction.}
\vspace{-2mm}
\begin{tabular}{|c|c|c|}
\hline
\multicolumn{1}{|c|}{\bf Model} & \multicolumn{1}{c|}{\bf Features used} & {\bf F1-score} \\ \hline
\multirow{2}{*}{\framework-24} & Content & 0.55 \\
 & Social & 0.58 \\ \hline
\multirow{2}{*}{R\&A} & Content & 0.52 \\
 & Social & 0.55 \\ \hline
\end{tabular}
\label{tab:non-temporal-feature-study}
\vspace{-5mm}
\end{table}
\subsubsection{Feature Ablation for Non-temporal Engagement}
For the non-temporal engagement prediction task, we perform feature ablation study for both \framework\ and R\&A.  Rowe and Alani \cite{rowe2014mining} grouped the features into two categories -- content features and social features. Content features of R\&A are closely similar to content, surface and latent features of \framework\ (see Sec. \ref{sec:feature_selection}), with many features common in both the models. We group these features as content features as a whole in this study, and name the user features as social features, for a better comparison between these two models. We use the best performing version of our model (\framework-24) for the feature importance study. Table \ref{tab:non-temporal-feature-study} shows that both the models treat social features with higher importance compared to content features for this task. 

\subsection{Performance of \embedding}
We compare \embedding\ with three baselines: Node2Vec \cite{Grover:2016} is run on (i) our co-occurrence matrix $\mathbf{A}$ ({\bf Node2Vec+$A$}) and (ii) a user-user matrix $\mathbf{A'}$, where $A'_{i,j}$ indicates the number of discussions where users $i$ and $j$ participated together ({\bf Node2Vec+$A'$}); third baseline is designed by aggregating all the comments/posts by user $i$ in the training set and running Doc2Vec \cite{le2014distributed} on the aggregated text to obtain user embedding ({\bf Doc2Vec})\footnote{The results of Node2Vec and Doc2Vec were reported after appropriate parameter tuning.}. Fig. \ref{fig:parameter}(c) shows that \embedding\ performs the best in both the tasks.  

We also present an empirical study on the complexity of \embedding. Intuitively, building the user-user co-occurrence matrix $\mathbf{A}$ is computationally most expensive, as it needs pair-wise comparison between users. Any pair-wise computation from an input of size $n$ results in an worst case time complexity of $\mathcal{O}(n^2)$. As we compute the full matrix, this bound should be same for space complexity, too. However, \embedding\ does not take pairs from the full user set $\mathbf{U}$ but only a finite subset $\mathbf{U}'\subset\mathbf{U}$. Complexity of building $\mathbf{A}$ is readily reflected by the number of non-zero elements in $\mathbf{A}$, because only these elements correspond to a pair-wise comparison between users. 
\begin{figure}
    \centering
    \includegraphics[width=0.75\columnwidth]{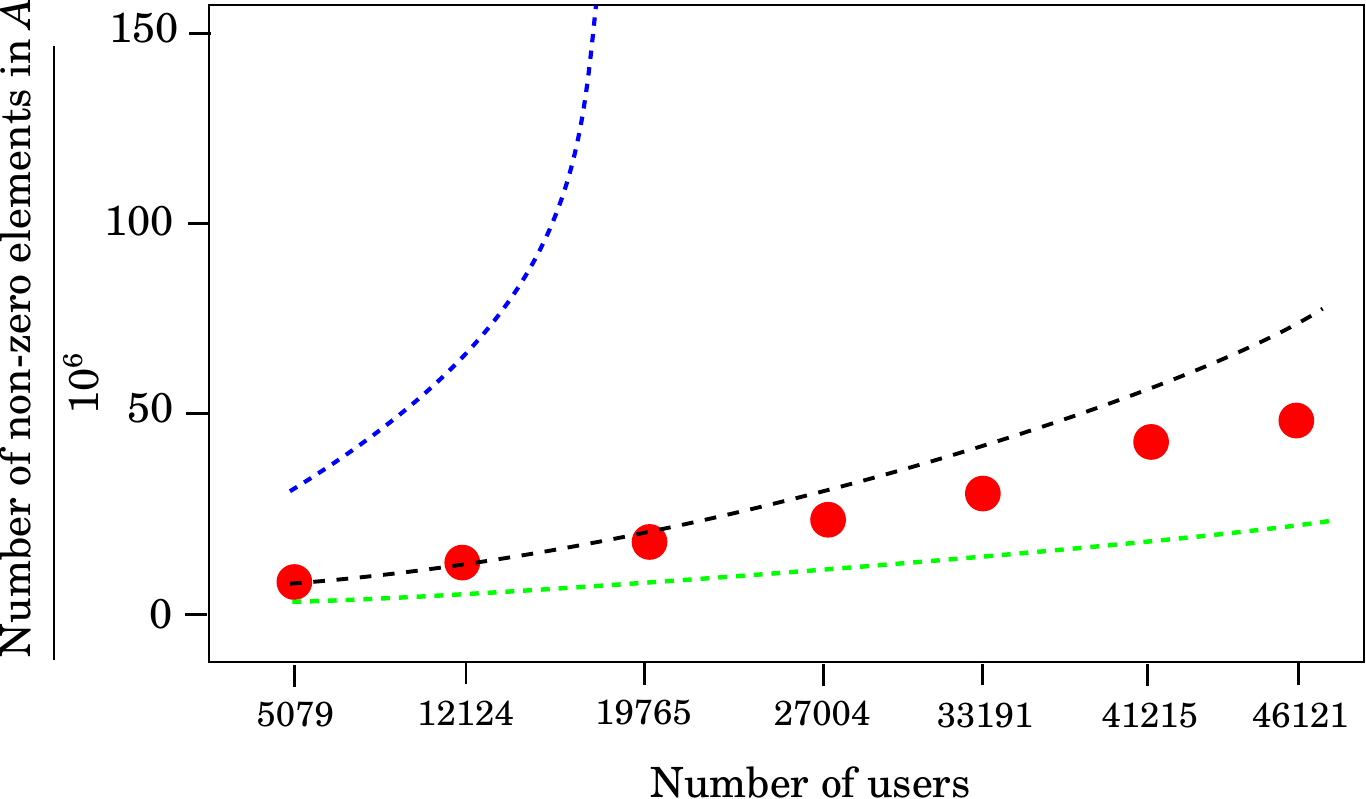}
    \vspace{-2mm}
    \caption{(Color online) Plot showing the number of non-zero elements in $\mathbf{A}$ versus total number of users (red dots); blue, black and green dashed lines represent the curves of $\lvert\mathbf{U}\rvert^2$, $\lvert\mathbf{U}\rvert^{\frac{5}{3}}$ and $\lvert\mathbf{U}\rvert^{\frac{3}{2}}$ respectively.}
    \vspace{-5mm}
    \label{fig:guvec_complexity}
\end{figure}
In Fig. \ref{fig:guvec_complexity}, we plot the number of non-zero elements in the co-occurrence matrix with varying sizes of user sets. For a comparative understanding, we also plot the curves of $\lvert\mathbf{U}\rvert^2$, $\lvert\mathbf{U}\rvert^{\frac{5}{3}}$ and $\lvert\mathbf{U}\rvert^{\frac{3}{2}}$. As we can see, the complexity of \embedding\ falls in between $\mathcal{O}(\lvert\mathbf{U}\rvert^{\frac{5}{3}})$ and $\mathcal{O}(\lvert\mathbf{U}\rvert^{\frac{3}{2}})$. This is due to high clustering of users in the user-user interaction network; users tend to form groups and their interactions remain mostly within the group. For this, \embedding\ needs to compute pair-wise proximity values for pairs from very small subsets of $\mathbf{U}$. Let us assume that the user set $\mathbf{U}$ is fragmented into $k$ equal-size partitions. Then, the total number of pair-wise computations \embedding\ needs will be $\frac{\lvert\mathbf{U}\rvert(\lvert\mathbf{U}\rvert-1)}{2k}$. From Fig. \ref{fig:guvec_complexity}, we know that,
$\lvert\mathbf{U}\rvert^{\frac{3}{2}} < \frac{\lvert\mathbf{U}\rvert(\lvert\mathbf{U}\rvert-1)}{2k} < \lvert\mathbf{U}\rvert^{\frac{5}{3}}$,
which further simplifies to bounds of $k$ itself, given by, 
$\mathcal{O}(\lvert\mathbf{U}\rvert^{\frac{1}{3}}) < k < \mathcal{O}(\lvert\mathbf{U}\rvert^{\frac{1}{2}})$.
\subsection{Diagnostics with a Case Study}
\begin{figure}[!t]
\centering
\includegraphics[width=0.5\textwidth]{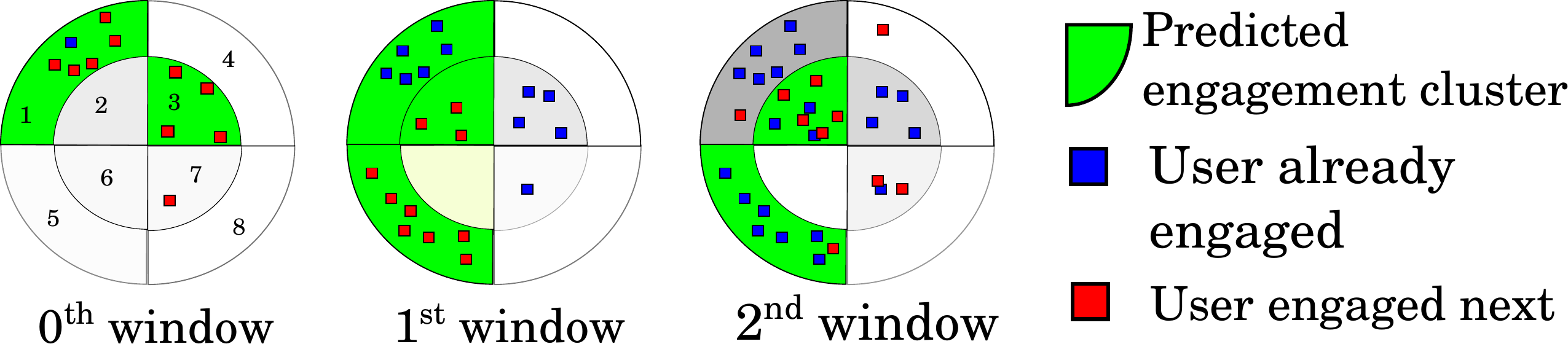}
\caption{(Color online) An example prediction of user clusters by \framework\ for first three consecutive windows. Eight clusters are marked by the indices in $0^{th}$ window.  Darker shades represent clusters with higher curvature value.}
\label{fig:example_engagement}
\vspace{-5mm}
\end{figure}

Fig.~\ref{fig:example_engagement} presents an example of  the user cluster engagement prediction results by \framework\ for first three consecutive windows.  We observe that \framework\ always computes high curvature for the cluster containing the users who started the discussion. It thus leads to an erroneous prediction for $1^{st}$ window, where \framework\ predicts that users from cluster-$1$ will be engaged. Even in the $2^{nd}$ window, a high curvature value is assigned to this cluster (darkest shade compared to rest of the clusters). Moreover, clusters from which users have been engaged in a window, tend to hold a high curvature value in the successive steps (cluster-$3$, for example). It is important to note that these are absolute values of curvature; originally, more attraction means more negative curvature.

For every window, \framework\ computes $g^{\mu\nu}$ for each cluster center. Using Eq.~\ref{Eq:metric_tensor}, we then compute intra-cluster distance for each cluster at every window. Table~\ref{tab:cluster_distance} shows that {\em metric distance} (distance between two vectors computed using Eq.~\ref{Eq:metric_tensor}) is always greater than flat Euclidean distance; more the curvature for a cluster (hence more probable the users from that cluster are to get engaged), more is the stretching of the intra-cluster distance.

 \section{Related Work}
Various social media platforms enable users with different types of activities.  In case of Twitter, a large body of literature address the problem of retweet prediction and user influence detection \cite{zhang2016retweet,cheng2014can}. Liu et al. \cite{liu2018c} proposed a user behavior model for retweet prediction. Recently, studies on the role of multimodality in retweet prediction have gained much focus \cite{wang2018retweet, zhao2018attentional}.
Another problem, which is much similar to ours, is the reply prediction \cite{nishi2016reply,yuan2016will,schantl2013utility,rowe2014mining}.
For both the tasks, various sets of features were employed, which  can be broadly categorized into two groups -- {\em content features}
and {\em social features}.
As Cha et al. \cite{cha2010measuring} and Macskassy and Michelson  \cite{macskassy2011people} suggested, content features play a vital role in retweet prediction; whereas replies are more dominated by social features \cite{sousa2010characterization, schantl2013utility}. Reply networks were studied in various other platforms: \url{Boards.ie}, SAP community network\footnote{https://www.sap.com/index.html}, Facebook and many others \cite{rowe2014mining}. Most of these studies  predict which post is going to get more replies, typically ignoring the temporal dynamics of discussions. Other studies explored the evolving structural properties of reply network  \cite{bakshy2011everyone,banos2013role,cogan2012reconstruction}. Purohit et al.
\cite{purohit2011understanding} proposed a framework to predict user engagement in clusters formed from topic-based discussions over Twitter.

\begin{table}[!t]
\centering
\caption{Average intra-cluster distances (FD represents Euclidean distances, MD-$i$ represents metric distance at $i^{th}$ window).}
\vspace{-3mm}
\scalebox{0.8}{
\begin{tabular}{|c|c|c c c|}
\hline
{\bf Cluster ID} & {\bf ED} & {\bf MD-0} & {\bf MD-1} & {\bf MD-2}\\
\hline
1 &0.259 &0.318 &0.323 &0.298\\
2 &0.179 &0.222 &0.226 &0.239\\
3 &0.210 &0.254 &0.247 &0.245\\
4 &0.286 &0.298 &0.301 &0.311\\
5 &0.232 &0.261 &0.275 &0.279\\
6 &0.175 &0.185 &0.181 &0.192\\
7 &0.198 &0.220 &0.223 &0.267\\
8 &0.205 &0.237 &0.229 &0.229\\
\hline
\end{tabular}}
\vspace{-5mm}

\label{tab:cluster_distance}

\end{table}
User-user engagement dynamics is a much studied problem, where the target is to predict the probability of future interaction between a pair of users based on their friendship history \cite{schantl2013utility, yuan2016will}. This is closely similar to link prediction in dynamic social networks \cite{ zhu2016scalable}.

Another related problem is comment popularity prediction in discussion forums. He et al. \cite{he2016deep} proposed a deep reinforcement learning model for predicting popular comments in Reddit. Horne et al. \cite{horne2017identifying} reported sentiment features to be most effective for comment popularity ranking in Reddit.

\section{Conclusion}
In this work, we adopted General Theory of Relativity to devise efficient fusion of heterogeneous features for modeling temporal and non-temporal dynamics of user engagement in online discussions. Our contributions in this work are: (i) \embedding, a novel user embedding method to represent users in a discussion platform as distributed vectors based on three different notions of proximity, (ii) \framework, a novel user engagement model inspired by Einstein Field Equations, (iii) a comprehensive set of features characterizing a discussion (post, comments, and users), and (iv) an exhaustive comparative analysis to show the superior performance of \framework\ compared to other baselines for temporal and non-temporal user engagement prediction and growth rate forecasting.

\bibliography{ref}
\bibliographystyle{IEEEtran}

\end{document}